%% file: paper.tex
\documentclass[oldversion]
{aa}

\usepackage{amsmath}
\usepackage{natbib}
\usepackage{epsfig}
\usepackage{txfonts}

\input{Befehle}

\newcommand{\change}[1]{{#1}}

\newcommand{\nbb}{{n_{\rm pl}}}
\newcommand{\nback}{{n_{\rm back}}}
\newcommand{\nbackA}{{A_{\rm back}}}

\newcommand{\Tb}{{T_{\rm b}}}
\newcommand{\ye}{{y_{\rm e}}}

\begin{document}


\titlerunning{Evolution of low-frequency features in the CMB spectrum}

\title{Evolution of low-frequency features in the CMB spectrum due to stimulated
  Compton scattering and Doppler-broadening}

\author{J. Chluba\inst{1} \and R.A. Sunyaev\inst{1,2}}
\authorrunning{Chluba \and Sunyaev}

\institute{Max-Planck-Institut f\"ur Astrophysik, Karl-Schwarzschild-Str. 1,
85741 Garching bei M\"unchen, Germany 
\and 
Space Research Institute, Russian Academy of Sciences, Profsoyuznaya 84/32,
117997 Moscow, Russia
}

\offprints{J. Chluba, 
\\ \email{jchluba@mpa-garching.mpg.de}
}

\date{Received / Accepted}

\abstract{We discuss a new solution of the Kompaneets-equation for physical
  situations in which low frequency photons, \change{forming relatively narrow
    spectral details, are Compton scattered} in an isotropic, infinite medium
  with an intense ambient blackbody field that is very close to full
  thermodynamic equilibrium with the free electrons.
  In this situation the {\it background-induced stimulated Compton scattering}
  slows down the motion of photons toward higher frequencies by a factor of
  $3$ in comparison with the solution that only takes into account
  Doppler-broadening and boosting.
  This new solution is important for detailed computations of cosmic microwave
  background spectral distortions arising due to uncompensated atomic
  transitions of hydrogen and helium in the early Universe.
In addition we derive another analytic solution that only includes the
background-induced stimulated Compton scattering and is valid for power-law
ambient radiation fields.
  This solution might have interesting applications for radio lines arising
  inside of bright extra-galactic radio source,
where according to our estimates line shifts because of background-induced
stimulated scattering could be amplified and even exceed the line broadening
due to the Doppler-effect.
}
\keywords{Cosmology: theory -- Cosmic Microwave Background: spectral distortions -- Compton scattering} 

\maketitle

\section{Introduction}
\label{sec:Intro}
The recombination of \ion{He}{III} ($4500 \lesssim z\lesssim 7000$) in the
Universe produces specific quasi-periodic distortions of cosmic microwave
background (CMB) in the Rayleigh-Jeans part of the CMB spectrum, at dm and cm
wavelength \citep{Dubrovich1997,Jose2007}.
It is also well-known that any energy release below $z\sim \pot{5}{4}$ will
lead to a characteristic broad $y$-type distortion \citep{Zeldovich1969} of
the background radiation.
In this case because of uncompensated loops \citep{Lyubarsky1983}, trying to
restore full equilibrium, similar features should originate from even higher
redshifts \citep{Chluba2008ydist}.
However, their amplitude and phase-dependence should differ from those of the
distortions arising during \ion{He}{III} recombination.

It is clear that \change{in both cases} the multiple scattering of photons by
hot electrons should lead to broadening and shifting of these spectral
features.
\change{Here} we investigate the evolution of narrow lines due to both {\it
  stimulated scattering} in the presence of the much brighter CMB background
with temperature $\Tg=2.7\,(1+z)\,$K, and {\it Doppler-broadening} and {\it
  boosting}.
According to our analytical solutions, in this situation the
background-induced stimulated Compton scattering at low frequencies slows down
the motion of photons toward higher frequencies by a factor of $3$ in
comparison with the solution that only takes into account Doppler-broadening
and boosting \citep{Zeldovich1969}.

We present two analytical solutions demonstrating this fact. In the first we
only include the pure effect of stimulated scattering in the presence of a
bright power-law ambient photon field, neglecting the Doppler-effect (see
Sect.~\ref{sec:Derivation_add}), while in the second solution we take into
account both effects simultaneously, but restrict ourselves to the case of a
low-frequency blackbody ambient photon field with temperature $\Tg\sim\Te$,
where $\Te$ denotes the electron temperature (see Sect.~\ref{sec:Derivation}).
The second solution also clearly shows that the broadening of the weak lines
in this situation depends only on the $y$-parameter defined by $\Te$, even
though the evolution of the ambient \change{CMB blackbody} spectrum itself is
described by $y\propto \Te-\Tg$.

\section{Summary of previous analytic solutions to the Kompaneets-Equation}
\label{sec:oldSols}
The repeated Compton scattering of photons by thermal electrons in isotropic,
infinite media can be described using the well-known Kompaneets-equation
\citep{Kompaneets1956}
\beal
\label{eq:Komp_eq}
\pAb{n(\xe, \ye)}{\ye}=\frac{1}{x_{\rm e}^2}\,\pAb{}{x_{\rm e}} \, x_{\rm e}^4 \left[\pAb{n}{x_{\rm e}}+n(1+n)\right]
\Abst{,}
\end{align}
where $n(\xe, \ye)=c^2 I_\nu/ 2 h\nu^3$ is the photon occupation number,
$\ye=\int\frac{k\Te}{\me c^2}\, \Ne\sigT c\id t$ is the Compton $y$-parameter,
where $\Te$ denotes the electron temperature, and $\xe=h\nu/k\Te$ is the
dimensionless frequency.
As is well understood, the first term in brackets ($\propto \partial_{x_{\rm
e}} n$) describes the {\it diffusion} of photons along the frequency axis, the
second term ($\propto n$) the motion of photons towards low frequencies due to
the {\it recoil}-effect, and the last term ($\propto n^2$) the effect of {\it
stimulated scattering}, which physically is also related to recoil
\citep[e.g. see][]{Sazonov2000}.
Eq.~\eqref{eq:Komp_eq} has been studied in great detail, both numerically
\citep[e.g.][]{Pozdniakov1983} and analytically in several limiting cases,
but, to our knowledge, no general analytic solution was found.

In situations when the recoil-term and stimulated scatterings are not
important (i.e. $n[n+1]\ll \partial_{x_{\rm e}} n$), \citet{Zeldovich1969} gave a
solution for arbitrary value of $\ye$, which reads
\beal
\label{eq:Sol_ZS}
\left.n(\nu, \ye)\right|_{\rm diff}=\frac{1}{\sqrt{4\pi\,\ye}}\int n_0(\tilde{\nu})
\,e^{-\textstyle\frac{(\ln[\nu/\tilde{\nu}]+3\ye)^2}{4\,\ye}}\,\frac{{\rm d} \tilde{\nu}}{\tilde{\nu}}
\Abst{.}
\end{align}
Here $n_0(\nu)$ is the photon occupation number at frequency $\nu$ and
$\ye=0$.
For $\ye\ll 1$ the broadening of an initially narrow line is given by
$\Delta\nu/\nu\sim \pm 2\sqrt{\ye\,\ln 2}$. Due to Doppler-boosting, the
maximum of the specific intensity $I_\nu$ moves along $\nu_{\rm
line}(\ye)=\nu_0\,e^{3\ye}$, so that for $\ye\ll 1$ one has $\Delta\nu/\nu\sim
3\ye$, implying that photons are upscattered.

In the case when the diffusion term and stimulated scatterings are not
important, \citet{Arons1971} and independently \citet{Illarionov1972}, using
different mathematical approach, gave a solution which can be
written in the form
\beal
\label{eq:Sol_IS}
\left.n(\nu, \tau_{\rm T})\right|_{\rm recoil}
=\frac{1}{[1-\omega\,\tau_{\rm T}]^4}\,n_0\left(\frac{\nu}{1-\omega\,\tau_{\rm T}}\right)
\Abst{,}
\end{align}
where $\omega=\frac{h\nu}{\me c^2}$ and $\tau_{\rm T}=\Ne\sigT c\,t$ is the
Thomson-scattering optical depth.
This solution simply describes the motion of photons towards lower
frequencies, where for initial photon distribution,
$n_0(\nu)=A\,\delta(\nu-\nu_0)/\nu^2$, at later time the line is located at
$\nu_{\rm line}(\tau_{\rm T})=\nu_0/[1+\omega_0 \tau_{\rm T}]$.
For $\omega\tau_{\rm T}\ll 1$ the line shift due to recoil is given by
$\Delta\nu/\nu\sim - \omega \tau_{\rm T}$.  This shows that at \change{low
  frequencies ($h\nu\ll k\Te$)} the recoil-effect can be neglected in comparison with
Doppler-boosting.
  
  The third analytic solution was found in the case when only the stimulated
  scattering term is important \citep{Zeldovich1968, Sunyaev1970induced}. Here
  the solution is determined by the implicit equation
\beal
\label{eq:Sol_S_induced}
\nu=\phi(f)-2\,\frac{h}{\me c^2}\,\tau_{\rm T}\,f
\Abst{,}
\end{align}
with $f(\nu,\tau_{\rm T})=\nu^2\,n(\nu,\tau_{\rm T})$ and where $\phi(z)$ can
be found from the initial condition ($\phi(z)\equiv f^{-1}_0(z)$, where
$f^{-1}_0(z)$ is the inverse function of $f(\nu,\tau_{\rm T})$ at $\tau_{\rm
  T}=0$).

Note that the last two solutions do not depend on the temperature of the
electrons, since in both cases physically the motion of the electrons in
lowest order does not play any role.

\section{Solution for purely background-induced stimulated scattering}
\label{sec:Derivation_add}
In this Section we discuss the situation when photons from a low-frequency
spectral feature are scattering off thermal electrons at temperature $\Te$
within an intense photon background that has a power-law spectral dependence,
i.e.  $\nback(\nu)=\nbackA\,\nu^{\beta}\gg 1$. Here $n_{0, \rm
back}=\nbackA\nu_0^\beta \gg 1$ is the occupation number of background photon
field in the vicinity of the spectral feature at $\nu_0$, and $\beta$ denotes
the power-law index\footnote{When the spectral intensity scales like
$I_\nu\propto \nu^{-\alpha}$, one has $\beta=-\alpha-3$. For the
Rayleigh-Jeans part of a blackbody spectrum $\beta=-1$.}.

If we neglect the evolution of the background photon field\footnote{This
assumption is for example justified when the frequency shifts of the
background radiation field are relatively small.} and assume that recoil and
Doppler broadening are negligible, then, inserting $n=\nback+\Delta n$ into
Eq.~\eqref{eq:Komp_eq}, one is left with
\beal
\label{eq:Komp_eq_D_add}
\pAb{\Delta n(\nu, \tau_{\rm T})}{\tau_{\rm T}}\!\approx\! 
\frac{2}{\nu^2}\,\frac{h}{\me c^2}\,\pAb{}{\nu} \, \nu^4\,\nback(\nu)\, \Delta n
\Abst{.}
\end{align}
%
For initial photon distribution $\Delta n_0(\nu)=\Delta n(\nu, \tau_{\rm
  T}=0)$ the solution is given by
\beal
\label{eq:Sol_Komp_eq_D_add}
\left.\Delta n(\nu, \tau_{\rm T})\right|_{\rm back-ind}
=
\frac{\Delta n_0\left(\frac{\nu}{[1-2(\beta+1)\,\nback(\nu)\,\omega\,\tau_{\rm
        T}]^{\frac{1}{\beta+1}}}\right)}
{[1-2(\beta+1)\,\nback(\nu)\,\omega\,\tau_{\rm T}]^{\frac{\beta+4}{\beta+1}}}
\Abst{.}
\end{align}
This solution describes the purely background-induced motion of photons along
the frequency axis. Due to the large factor $2\,\nback\gg 1$ the speed of this
motion is increased, so that even for $\omega\tau_{\rm T}\ll 1$ the line shift
can still be considerable.
For $\beta >-1$ this solution is valid only at
$\nu<1/[2(\beta+1)\,\nbackA\,\frac{h \tau_{\rm T}}{\me c^2}]^{1/(\beta+1)}$.

For initial photon distribution, $\Delta n_0(\nu)=A\,\delta(\nu-\nu_0)/\nu^2$,
the line is located at $\nu^{\delta}(\tau_{\rm
T})=\nu_{0}/[1+2(\beta+1)\,\nback(\nu_0)\,\omega_0\,\tau_{\rm T}]^{\frac{1}{\beta+1}}$.
Here the solution works for all $\beta >-1$, while for $\beta <-1$ it is valid
only for $\nu_0>[2|\beta+1|\,\nbackA\,\frac{h \tau_{\rm T}}{\me c^2}]^{1/|\beta+1|}$.
For the condition $2(\beta+1)\,\nback(\nu_0)\,\omega_0\,\tau_{\rm T}\ll 1$ one
has $\Delta\nu/\nu_0\sim -2\,\nback(\nu_0)\,\omega_0\,\tau_{\rm T}$. This shows
that independent of $\beta$ photons are {\it always} moving toward lower
frequencies, where the speed of this motion is increased by $2\,\nback\gg 1$
as compared to the recoil-case (see Sect.~\ref{sec:oldSols}).
It is also remarkable that for $\beta=-1$, \change{i.e. $\nback\propto
1/\nu$,} one has $\Delta\nu/\nu\approx -2\,\nbackA \frac{h \tau_{\rm T}}{\me c^2}$, so that the
line shift is independent of frequency.

It is also possible to rewrite this expression in terms of the brightness
temperature 
\beal
\label{eq:Tb}
T_{\rm b}(\nu)=\frac{c^2 I_\nu^{\rm back}}{2 k\nu^2}\equiv \frac{h\nu}{k}\,\nback(\nu), 
\end{align}
of the background field in the vicinity of the spectral feature, yielding
\beal
\label{eq:Dnu_nu_f}
\left.\frac{\Delta\nu}{\nu}\right|_{\rm back-ind} 
\sim -2\,\frac{k T_{\rm b}(\nu)}{\me c^2}\tau_{\rm T}.
\end{align}
%
For $T_{\rm b}>\frac{3}{2}\Te$, it is clear that the line shift due to
background-induced stimulated scattering exceeds the shift towards higher
frequencies due to Doppler-boosting. Therefore the net motion of the line
center can be directed towards lower frequencies, \change{even if one includes
  the Doppler-term}.
\change{Note that for a Rayleigh-Jeans spectrum $\Tb$ is equal to the physical
  temperature of the radiation field.}

\change{Comparing with the case of pure Doppler-broadening (see
  Sect.~\ref{sec:oldSols}),} it is also clear that for $\ye\ll 1$ and
$\ye\gtrsim 4\, [\Te/T_{\rm b}(\nu)]^2 \ln 2$ the background-induced line
shift becomes larger than the FWHM Doppler-broadening connected with the
diffusion term.
\change{In this case one can neglect the effect of line-broadening due to the
  Doppler-term, and, according to Eq.~\eqref{eq:Dnu_nu_f}, even for small
  Thomson optical depth can still obtain a rather significant line-shift.
  In radio spectroscopy is possible to determine tiny shifts or broadening in
  the frequency of narrow spectral feature (for example of 21 cm lines from
  high redshift radio galaxies).

}

\subsection{Rayleigh-Jeans limit ($\beta=-1$)}
For $\beta=-1$ Eq. \eqref{eq:Komp_eq_D_add} becomes $\partial_{\tau_{\rm T}}
\Delta n(\nu, \tau_{\rm
T})\!=\!\frac{2}{\nu^2}\,\frac{h\,\nbackA}{\me c^2}\,\partial_\nu \,
\nu^3\, \Delta n$, and with the replacement $s=\nu^3 \, \Delta n$ and
$\xi=\ln(\nu)$, can be cast in the form $\partial_{\tau_{\rm T}} s(\xi,
\tau_{\rm T})\!=\!\frac{2h\,\nbackA}{\me c^2}\,\partial_\xi s$.
Using the characteristic of this equation, or by directly taking the limit
$\beta\rightarrow -1$ from Eq.~\eqref{eq:Sol_Komp_eq_D_add}, one
readily finds
\beal
\label{eq:Sol_Komp_eq_D_add_m1}
\left.\Delta n(\nu, \tau_{\rm T})\right|^{\beta=-1}_{\rm back-ind}
=e^{6\nbackA\,\frac{h\,\tau_{\rm T}}{\me c^2}}\,\Delta n_0\left(\nu\,e^{2\nbackA\,\frac{h\,\tau_{\rm T}}{\me c^2}}\right)
\Abst{,}
\end{align}
so that for $\Delta n_0(\nu_0)=A\,\delta(\nu-\nu_{0})/\nu^2$, the line will be
centered at $\nu^{\delta}(\tau_{\rm
T})=\nu_0\,e^{-2\nbackA\,\frac{h\,\tau_{\rm T}}{\me c^2}}$.
If we assume that $\nback=k\Tg/h\nu$, i.e. we are in the Rayleigh-Jeans part
of a blackbody photon field with $\Tg=\Tb=\Te$, then we have
$\nu^{\delta}(\tau_{\rm T})=\nu_{0}\,e^{-2 \frac{k\Te}{\me c^2}\,\tau_{\rm
T}}$.
This shows that due to stimulated scattering of photons within a blackbody
ambient radiation field are moving with a speed $\propto 2 \ye$ towards lower
frequencies.
This is 3/2 times smaller than the line shift due to Doppler-boosting (see
Sect.~\ref{sec:oldSols}), but is directed in the opposite direction.
Below we will discuss this case in more detail, also including the broadening
of the line due to the Doppler-effect.

\subsection{Evolution of the background-field and the relative
  velocity of a spectral feature for small line-shifts}
In the derivation given above we have neglected the evolution of the
background radiation field. This is possible when the line-shifts
$\Delta\nu/\nu$ are small.
If we insert $n=\nback$ into the Kompaneets-equation \eqref{eq:Komp_eq}, and
assume \change{that $\ye$ is sufficiently small}, then we can directly write
the solution for the change of the photon occupation number as
\beal
\label{eq:nline_smally}
\frac{\Delta \nback}{\nback}
&=\ye\left\{\beta(3+\beta)+(4+\beta)\xe+2\nback(2+\beta)\xe\right\}.
\end{align}
With this one can now find the corresponding overall shift of the background
radiation field at initial frequency $x_{\rm e, 0}$. Solving the equation
$\nback(x_{\rm e, 0})\equiv \nback(\xe)[1+\frac{\Delta
\nback(\xe)}{\nback(\xe)}]$, with $\xe=x_{\rm e, 0}(1+\delta)$, for
$\delta\equiv -\Delta\nu/\nu$ under the condition $\beta\neq 0$ one finds:
\beal
\label{eq:back_shift}
\left.\frac{\Delta\nu}{\nu}\right|_{\rm back}&\sim 
\frac{2\,(2+\beta)}{\beta}\frac{k T_{\rm b}(\nu)}{\me c^2}\tau_{\rm T}
+ \frac{4}{\beta}\,\frac{h\nu}{\me c^2}\,\tau_{\rm T}
\nonumber\\
&\qquad\quad
+(3+\beta+\xe)\,\frac{k \Te}{\me c^2}\tau_{\rm T} 
\Abst{.}
\end{align}
%
With $|\Delta\nu/\nu|\ll 1$ one can give the condition under which
it is possible to neglect the evolution of the background field.

\change{It is important to mention that Eq.~\eqref{eq:nline_smally} does only
  conserve a Wien spectrum ($n=e^{-\xe}\approx 1$ for $\xe\ll 1$) and
  Rayleigh-Jeans spectrum ($n=1/\xe$) to order $\xe$. This is because higher
  order terms cannot be taken into account by simple power-laws. Only
  $n=1/[e^{\xe+\mu}-1]$ with constant $\mu$ and $\Te=\Tg$ is truly
  invariant. }

\section{Scattering of low frequency photons by free electrons in an intense ambient blackbody field}
\label{sec:Derivation}
If we assume that at $\ye=0$ the initial photon field is given by
$n=\nbb(x_{\rm e})+\Delta n$, where $\nbb(x)=1/[e^x-1]$ is the blackbody
occupation number, $x_{\rm e}=h\nu/k\Te$, and $\Delta n$ is a small ($\Delta
n/\nbb(x_{\rm e})\ll 1$) spectral distortion, then with $\partial_{\xe}
\nbb(x_{\rm e})\equiv-\nbb(x_{\rm e})[1+\nbb(x_{\rm e})]$ and $\partial_\ye
\nbb(x_{\rm e})=0$, from Eq.~\eqref{eq:Komp_eq} one has
\beal
\label{eq:Komp_eq_D}
\pAb{\Delta n(\xe, \ye)}{\ye}\!=\!\frac{1}{x_{\rm e}^2}\,\pAb{}{\xe} \, x_{\rm
e}^4 \left[\pAb{\Delta n}{\xe}+\Delta n(1+2\nbb(x_{\rm e}) + \Delta n)\right]
\Abst{.}
\end{align}
If we now neglect terms of $\mathcal{O}(\Delta n ^2)$ and assume that $\xe\ll
1$, and hence $\nbb(x_{\rm e})\approx 1/x_{\rm e}\gg 1$, then
Eq.~\eqref{eq:Komp_eq_D} reads
\beal
\label{eq:Komp_eq_D_lin}
\pAb{\Delta n(\xe, \ye)}{\ye}=
\frac{1}{x_{\rm e}^2}\,\pAb{}{\xe} \, x_{\rm e}^4 \left[\pAb{\Delta n}{\xe}+ 2 \frac{\Delta n}{\xe}\right]
\Abst{.}
\end{align}
Introducing the variables $\xi=\ln(\xe)$ and $s=\xe^3 \Delta n$ we
then have 
%
$\partial_\ye s(\xi, \ye)=\partial_\xi^2 s-\partial_\xi s$.
%
Similar to \citet{Zeldovich1969} we now transform to $z=\xi-\ye$ so that
this equation 
%
%
reduces to the normal diffusion equation $\partial_\ye s=\partial_z^2 s$.
Therefore the solution of Eq.~\eqref{eq:Komp_eq_D_lin} is
\beal
\label{eq:Sol}
\Delta n(\nu, \ye)=\frac{1}{\sqrt{4\pi\,\ye}}\int \frac{\tilde{\nu}^3}{\nu^3}\,\Delta
n(\tilde{\nu}, 0)
\,e^{-\textstyle\frac{(\ln[\nu/\tilde{\nu}]-\ye)^2}{4\,\ye}}\,\frac{{\rm d} \tilde{\nu}}{\tilde{\nu}}
\Abst{.}
\end{align}
Not including the term $2 \frac{\Delta n}{\xe}$ in
Eq.~\eqref{eq:Komp_eq_D_lin}, simply leads to the replacement $\ye\rightarrow
3\ye$ in the exponential term of Eq.~\eqref{eq:Sol}, and after absorbing the
factor $\tilde{\nu}^3/\nu^3=\exp(-3\ln[\nu/\tilde{\nu}])$ one obtains the
solution of \citet{Zeldovich1969} in the form of Eq.~\eqref{eq:Sol_ZS}.

If we assume $\Delta n(\nu, 0)=A\,\delta(\nu-\nu_0)/\nu^2$ for the initial
distortion then from Eq.~\eqref{eq:Sol} we obtain
\beal
\label{eq:Sol_delta}
\Delta n(\nu, \ye)=\frac{A}{\sqrt{4\pi\,\ye}\,\nu^3}\times e^{-\textstyle\frac{(\ln[\nu/\nu_0]-\ye)^2}{4\,\ye}}
\Abst{.}
\end{align}
Without the inclusion of stimulated scattering in the ambient
blackbody field (replacement $\ye\rightarrow 3\ye$ in Eq.~\eqref{eq:Sol}), as in
the case of \citet{Zeldovich1969}, one finds
\bsub
\label{eq:Sol_delta_ZS}
\beal
\label{eq:Sol_delta_ZS_a}
\Delta n^{\rm ZS}(\nu, \ye)
&=\frac{A}{\sqrt{4\pi\,\ye}\,\nu^3}\times e^{-\textstyle\frac{(\ln[\nu/\nu_0]-3\,\ye)^2}{4\,\ye}}
\\
\label{eq:Sol_delta_ZS_b}
&\equiv \frac{A}{\sqrt{4\pi\,\ye}\,\nu_0^3}\times e^{-\textstyle\frac{(\ln[\nu/\nu_0]+3\,\ye)^2}{4\,\ye}}
\Abst{,}
\end{align}
\esub
Comparing Eq.~\eqref{eq:Sol_delta} and \eqref{eq:Sol_delta_ZS_a} one can see
that in terms of energy ($\Delta I_\nu\propto\nu^3 \Delta n$) the maximum of the
distribution is moving like $\nu_{\rm max}(\ye)=\nu_0\,e^\ye$, when including
the effect of stimulated scattering, while it moves like $\nu^{\rm ZS}_{\rm
max}(\ye)=\nu_0\,e^{3\ye}$ without this term.
On the other hand, in terms of photon number ($\Delta N_\nu\propto\nu^2 \Delta n$),
from Eq.~\eqref{eq:Sol_delta} and \eqref{eq:Sol_delta_ZS_a} one finds that the
maximum of the distribution is moving like $\nu_{\rm
max}(\ye)=\nu_0\,e^{-\ye}$, when including the effect of stimulated
scattering, while it moves like $\nu^{\rm ZS}_{\rm max}(\ye)=\nu_0\,e^{\ye}$
without this term.
In both cases the difference in the line position is a factor of $e^{-2 \ye}$,
a property that is in agreement with the more simple solution
\eqref{eq:Sol_Komp_eq_D_add_m1}.

\begin{figure}
\centering 
\includegraphics[width=0.9\columnwidth]{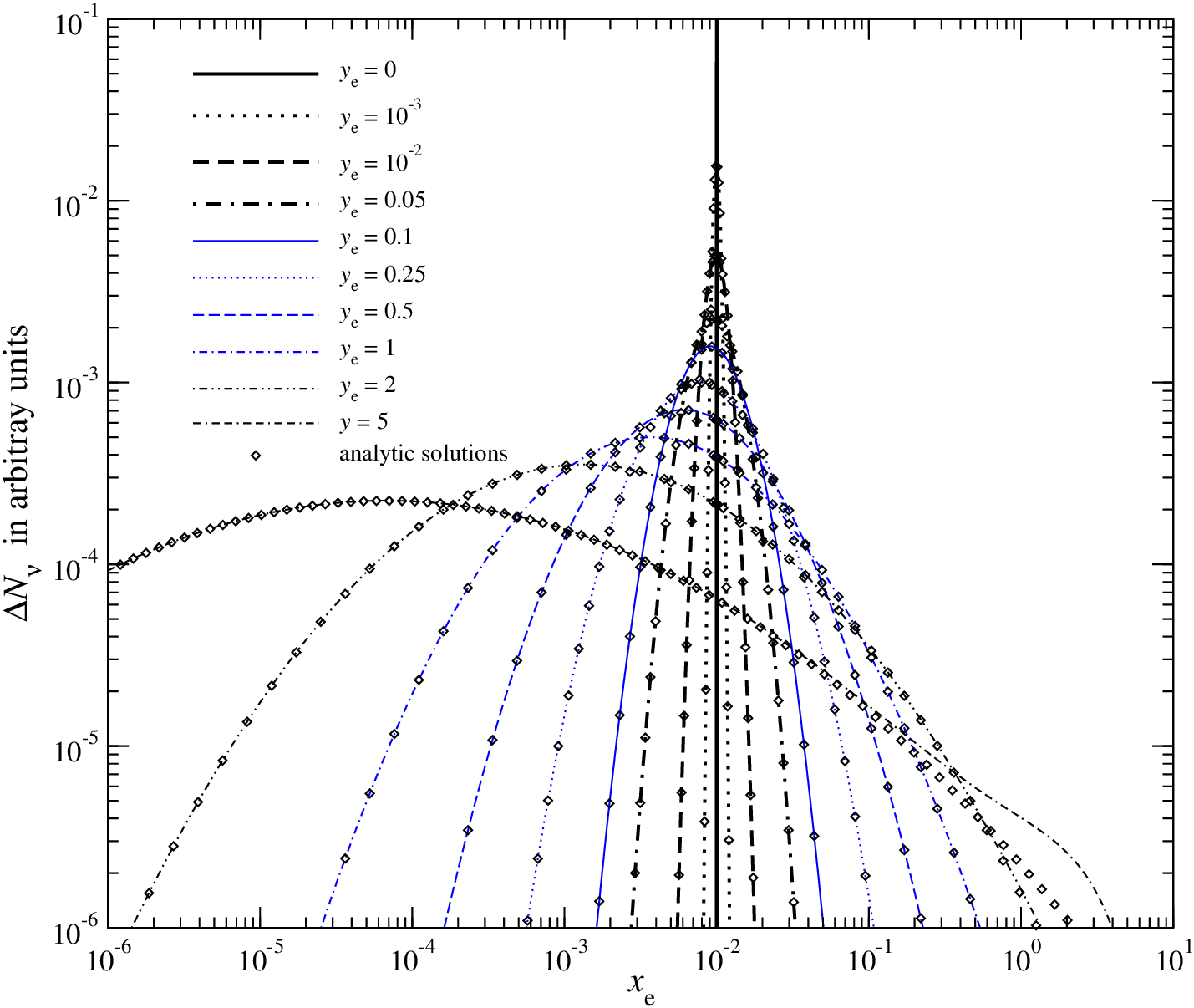}
  \\
\includegraphics[width=0.9\columnwidth]{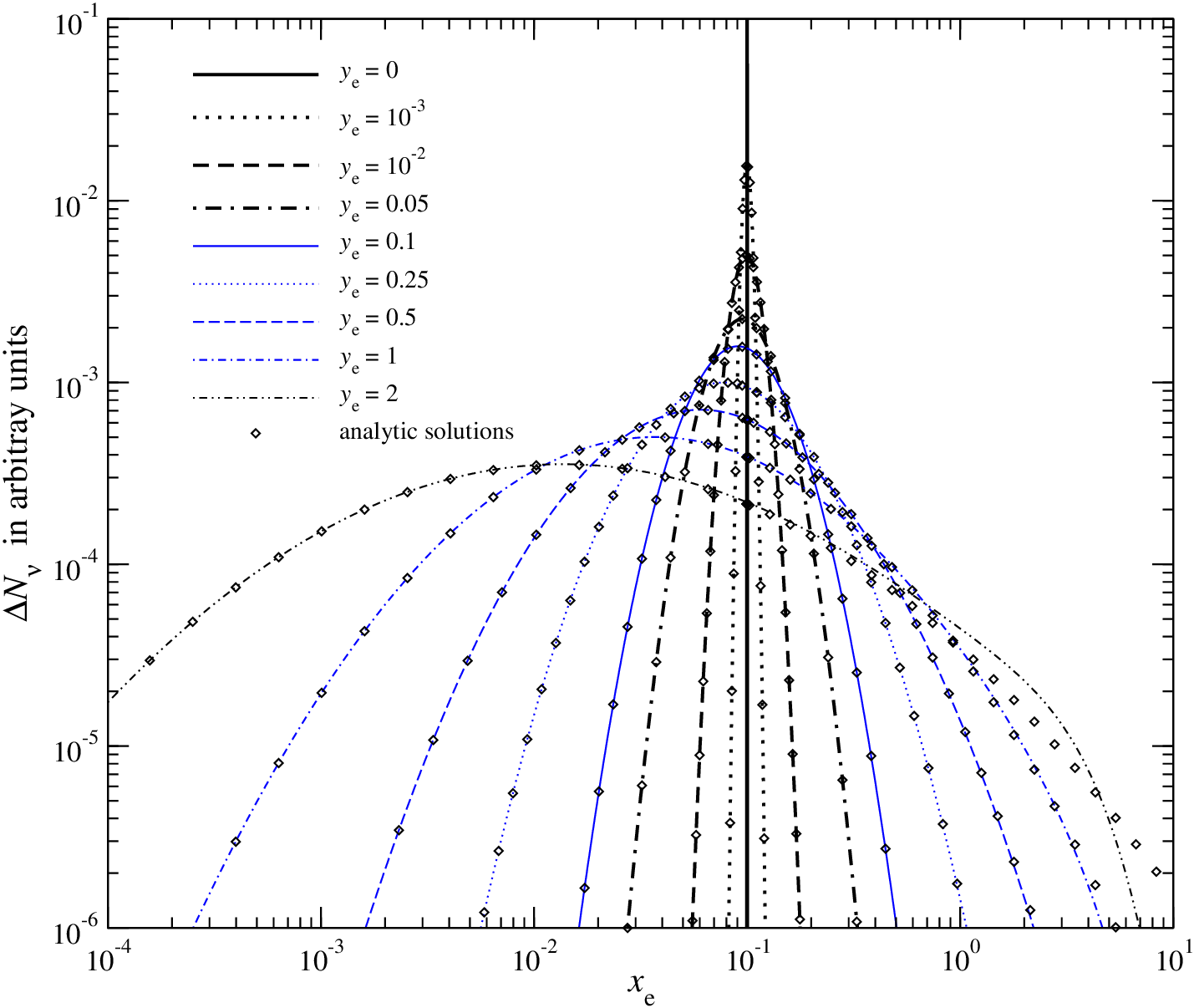}
\caption{Time-evolution of $\Delta N_\nu=\nu^2 \Delta n$ for different values
of the $y$-parameter {\it including} the effect of stimulated scattering in
the blackbody ambient radiation field.  The upper panel shows the case, for an
initially narrow line which was injected at frequency $x_{\rm e, 0}=10^{-2}$,
while the lower panel shows the solution for injection at $x_{\rm e,
0}=10^{-1}$. In both figures we show the results as obtained by numerically
solving Eq.~\eqref{eq:Komp_eq_D} for $\Tg=\Te$. In addition we give the
analytic solutions of the linearized problem, Eq.~\eqref{eq:Komp_eq_D_lin},
according to Eq.~\eqref{eq:Sol}.  }
\label{fig:Nx}
\end{figure}

\begin{figure}
\centering 
\includegraphics[width=0.9\columnwidth]{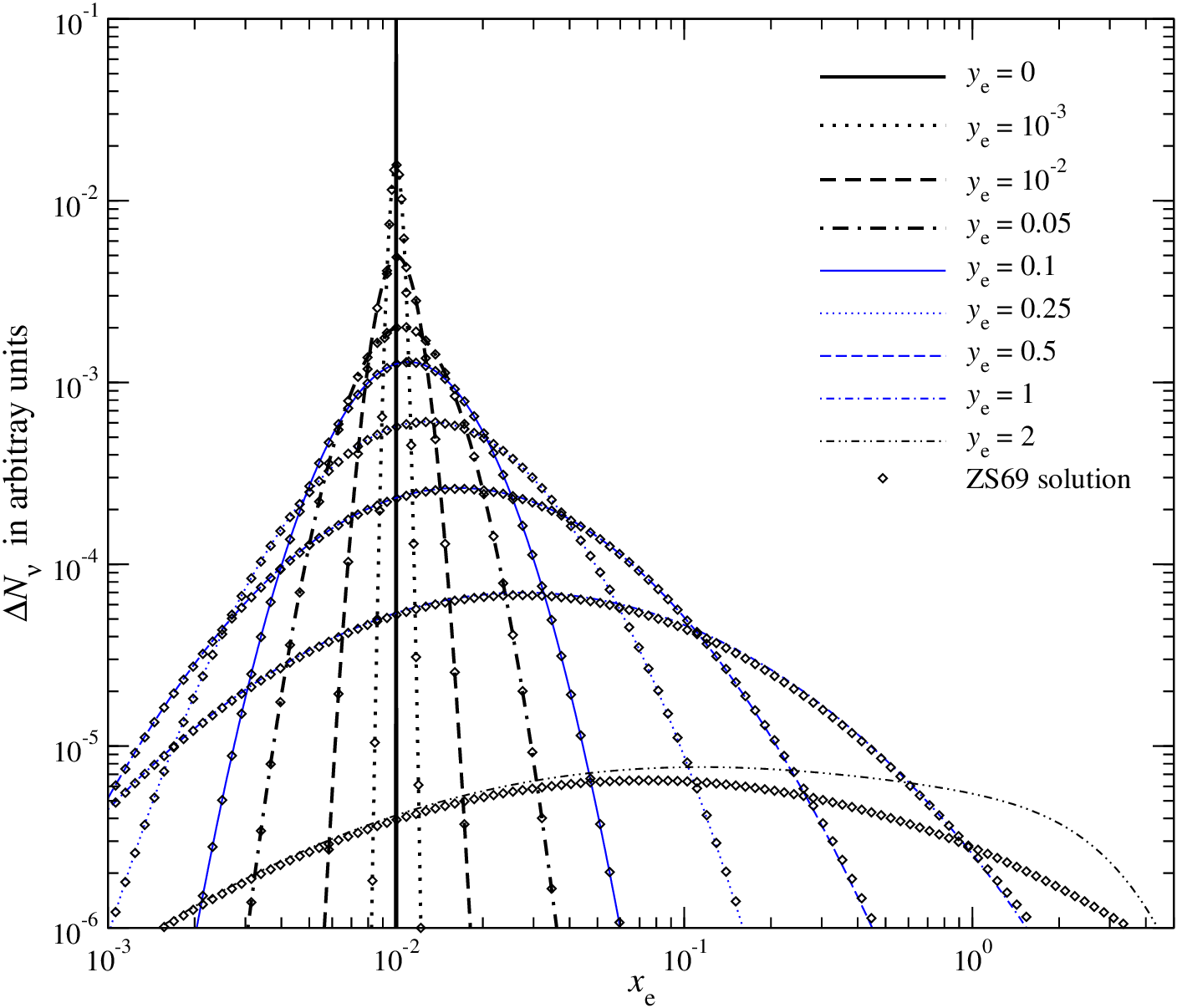}
\\
\includegraphics[width=0.9\columnwidth]{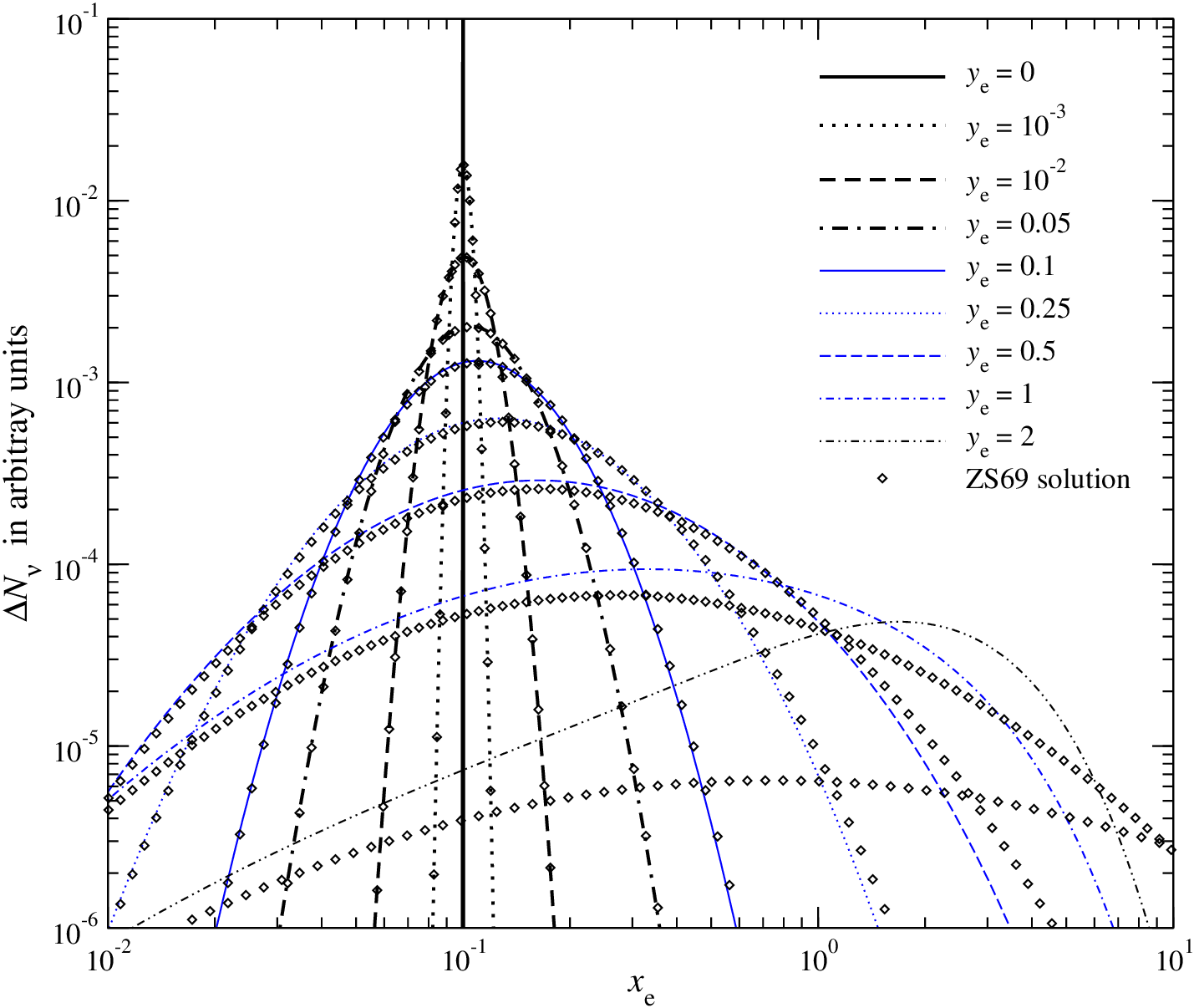}
\caption{Time-evolution of $\Delta N_\nu=\nu^2 \Delta n$ for different values
of the $y$-parameter {\it not including} the effect of stimulated scattering
in the blackbody ambient radiation field ($\nbb=0$).  The upper panel shows
the case, for an initially narrow line which was injected at $x_{\rm e,
0}=10^{-2}$, while the lower panel shows the solution for injection at
$x_{\rm e, 0}=10^{-1}$. In both figures we show the results as obtained by numerically
solving Eq.~\eqref{eq:Komp_eq_D} for $\Tg=\Te$. In addition we give the
analytic solutions according to \citet{Zeldovich1969}, Eq.~\eqref{eq:Sol_ZS}.
}
\label{fig:Nx_ZS}
\end{figure}
As an example, in Fig.~\ref{fig:Nx} and \ref{fig:Nx_ZS} we show the
time-evolution of an initially narrow line for different values of the
$y$-parameter. For the curves shown in Fig.~\ref{fig:Nx} the effect of
stimulated scattering in the blackbody ambient radiation field was included,
while in Fig~\ref{fig:Nx_ZS} it was not.
As expected, in the former case the maximum of the distribution is moving
toward lower frequencies, while in the latter it moves toward higher
frequencies.
It is also remarkable that the analytic solution \eqref{eq:Sol} in comparison
with the full numerical solution of Eq.~\eqref{eq:Komp_eq_D} works up to
values of $\ye\sim 1-5$. It only breaks down when the photons reach the region
$\xe\gtrsim 1$, where the recoil effect starts to become important. 
However, in both considered cases (injection at $x_{\rm e, 0}=10^{-2}$ and
$x_{\rm e, 0}=0.1$) only a small fraction of the initial number of photons
reaches this region for $\ye\lesssim 1$.
\change{One can also see that no matter if the background-induced effect is
  included or not, the line-broadening is not changing much. It is only the
  position of the line center that is affected by this process, but not the
  Doppler-broadening.}

\section{Discussion and conclusion}
\label{sec:Conc}
We have presented a new solution of the Kompaneets-equations for physical
situations in which low frequency photons are scattered by thermal electrons
but within an intense ambient blackbody field that is very close to full
thermodynamic equilibrium with the free electrons. 
Under these circumstances the blackbody-induced stimulated scattering of
photons slows down the motion of photons toward higher frequencies by a factor
of $3$ in comparison with the solution that only take into account
Doppler-broadening and boosting \citep{Zeldovich1969}.

These physical conditions are met during the epoch of cosmological
recombination, in particular for
$\ion{He}{iii}\rightarrow\ion{He}{ii}$-recombination, where electron
scattering can still affect the cosmological recombination spectrum notably
\citep{Dubrovich1997, Jose2007}.
The CMB blackbody spectrum is an exact solution of the Kompaneets equation when
$\Te=\Tg$. Nevertheless, the weak low frequency features in the presence of
CMB with $\Te=\Tg$ are evolving and boosting their central frequencies due to
the simultaneous action of induced scattering and Doppler-effect resulting
after each scattering. The additional line shift towards low frequencies due
to recoil effect is small because $\xe\ll 1$.

Since at $z\sim 6000$, i.e. where most of the \ion{He}{ii}-photons are
released, one finds $\ye\sim \pot{2}{-3}$, so that within the standard
recombination epoch this process is only important for very accurate
predictions of the cosmological recombination spectrum at low frequencies.
These may become necessary for the purpose of using the cosmological
recombination spectrum for measurements of cosmological parameters, such as
the CMB temperature monopole, the specific entropy of the Universe, or the
pre-stellar abundance of helium, as discussed recently \citep{RS2008}.
Our analysis shows that the low-frequency spectral features from
$\ion{He}{iii}\rightarrow\ion{He}{ii}$ recombination should then be shifted by
$\Delta\nu/\nu\sim 0.2\%$ instead of $\sim 0.6\%$.
At the same time the Doppler-broadening of these lines reaches $\sim 15\%$ at
FWHM, and the typical line width is $\Delta\nu/\nu\sim 20-30\%$
\citep{Jose2007}.

However, if photons due to uncompensated atomic transitions in hydrogen and
helium were released much earlier ($z\gtrsim 10^4-10^5$) then the
blackbody-induced stimulated scattering becomes more significant.
As mentioned in the introduction, such emission can occur in connection with
possible intrinsic spectral distortions of the CMB, due to energy release in
the early Universe \citep{Chluba2008ydist}.
%

We should stress that the discussed behavior of spectral features is
characteristic only for the low frequency part of CMB spectrum $h\nu\ll
k\Tg$. In the dm and cm spectral range this condition is fulfilled with very
good accuracy, and the occupation number of photons exceeds $\sim 10$. The
broadening and shifting of the photons in the Wien-part of the spectrum has
completely different behavior and will be discussed in a separate paper.

In addition we gave the solution \eqref{eq:Sol_Komp_eq_D_add} which describes
the evolution of narrow spectral features in the presence of a very bright,
broad-band radiation background with $\nback\gg 1$.
It might have interesting applications for radio lines arising inside of
extra-galactic radio source.
where according to our estimates (see Sect.~\ref{sec:Derivation_add}) the
background-induced line shifts could be amplified and even exceed the line
broadening due to the Doppler-effect.
%
%


\bibliographystyle{aa} 
\bibliography{Lit}

\end{document}

%% file: Befehle.tex
\newcommand{\xe}{x_{\rm e}}

\newcommand{\id}{{\,\rm d}}

\newcommand{\beq}{\begin{equation}}   %

\newcommand{\eeq}{\end{equation}}   %

\newcommand{\beqa}{\begin{eqnarray}}   %

\newcommand{\eeqa}{\end{eqnarray}}   %

\newcommand{\beal}{\begin{align}}

\newcommand{\enal}{\end{align}}

\newcommand{\bspl}{\begin{split}}

\newcommand{\espl}{\end{split}}

\newcommand{\bsub}{\begin{subequations}}

\newcommand{\esub}{\end{subequations}}

\newcommand{\bmulti}{\begin{multline}}   %

\newcommand{\beqm}{\begin{mathletters}}   %

\newcommand{\eeqm}{\end{mathletters}}   %

\newcommand{\Abst}[1]{\,#1}

\newcommand{\me}{m_{\rm e}}

\newcommand{\Ne}{N_{\rm e}}

\newcommand{\Te}{T_{\rm e}}

\newcommand{\Tg}{T_{\gamma}}

\newcommand{\sigT}{\sigma_{\rm T}}

\newcommand{\pd}{\partial}

\newcommand{\pAb}[2]{\frac{\displaystyle\pd #1}{\displaystyle\pd #2}}

\newcommand{\pot}[2]{#1 \times 10^{#2}}

